\documentclass[twocolumn,tighten]{aastex62}

\usepackage{apjfonts}
\usepackage{kotex}

\shorttitle{$\MakeLowercase{hk}$ observation of dSph's}
\shortauthors{Han et al.}

\begin{document}

\title{Narrowband Ca Photometry for Dwarf Spheroidal Galaxies. I. Chemostructural Study on Draco, Sextans, and Canes Venatici I\footnote{Based on data collected at the Subaru Telescope, which is operated by the National Astronomical Observatory of Japan.}}


\correspondingauthor{Suk-Jin Yoon; sjyoon0691@yonsei.ac.kr}

\author{Sang-Il Han}
\altaffiliation{Both authors have contributed equally to this paper.}
\affil{Department of Astronomy and Center for Galaxy Evolution Research, Yonsei University, Seoul 03722, Republic of Korea}
\affil{Korea Astronomy and Space Science Institute, 776 Daedeokdae-ro, Yuseong-gu, Daejeon, 34055, Republic of Korea}

\author[0000-0001-7033-4522]{Hak-Sub Kim}
\altaffiliation{Both authors have contributed equally to this paper.}
\affil{Korea Astronomy and Space Science Institute, 776 Daedeokdae-ro, Yuseong-gu, Daejeon, 34055, Republic of Korea}

\author[0000-0002-1842-4325]{Suk-Jin Yoon}
\affil{Department of Astronomy and Center for Galaxy Evolution Research, Yonsei University, Seoul 03722, Republic of Korea}

\author{Young-Wook Lee}
\affil{Department of Astronomy and Center for Galaxy Evolution Research, Yonsei University, Seoul 03722, Republic of Korea}

\author{Nobuo Arimoto}
\affil{Subaru Telescope, National Astronomical Observatory of Japan, 650 North A’ohoku Place, Hilo, HI 96720, USA}
\affil{Astronomy Program, Department of Physics and Astronomy, Seoul National University, Seoul 08826, Republic of Korea}
\affil{National Astronomical Observatory of Japan, 2-21-1 Osawa, Mitaka, Tokyo 181-8588, Japan}

\author{Sakurako Okamoto}
\affil{Subaru Telescope, National Astronomical Observatory of Japan, 650 North A’ohoku Place, Hilo, HI 96720, USA}
\affil{National Astronomical Observatory of Japan, 2-21-1 Osawa, Mitaka, Tokyo 181-8588, Japan}
\affil{The Graduate University for Advanced Studies, Osawa 2-21-1, Mitaka, Tokyo 181-8588, Japan}

\author{Chang H. Ree}
\affil{Korea Astronomy and Space Science Institute, 776 Daedeokdae-ro, Yuseong-gu, Daejeon, 34055, Republic of Korea}


\begin{abstract}

A few dozen dwarf satellite galaxies of the Milky Way have been discovered, which are often viewed as the remaining building blocks of our Galaxy. 
The follow-up spectroscopy showed that dwarf galaxies have a sizeable spread in their metallicities. 
Several scenarios were suggested to explain the metallicity spread, which can be tested by the structural patterns of stellar subpopulations with distinct metallicities. 
However, such chemical plus structural examination, to which we refer to as ``chemostructural study'' is hindered by the lack of stars with spectroscopic metallicity. 
Here we propose the Ca--$by$ photometry as an alternative way to secure metallicities for a 2--3 orders of magnitude larger stellar sample than the spectroscopic sample and thus enable us to perform a chemostructural study on dwarf galaxies. 
In particular, we use the $hk$ index [$\equiv($Ca$-b)-(b-y)$], whose validity as a photometric metallicity indicator (and crass insensitivity to age) for red-giant-branch stars was upheld via Galactic globular clusters, and observe three dwarf spheroidal galaxies---Draco, Sextans, and Canes Venatici I (CVnI)---with Subaru/Suprime-Cam. 
We find that in all the galaxies the metal-rich stellar populations are more centrally concentrated than the metal-poor counterparts, suggesting that the central regions of the galaxies underwent extended star formation.
Such a negative radial metallicity gradient for Sextans and CVnI opposes to the traditional spectroscopic results. 
We also find that their metallicity distribution functions (MDFs) can be characterized by a unimodal, skewed Gaussian shape with a metal-rich peak and a metal-poor tail.
We discuss their features in the chemostructure and MDF in terms of dwarf galaxy formation theories.

\end{abstract}


\section{Introduction} \label{sec:intro}

\begin{table*}
\caption{Summary of observations}
\begin{center}
\begin{tabular}{lcccccr}
\hline\hline

Galaxy  &  R.A.  &  Decl.  &  Filter  &  Exposure Time  &  Date  &  Seeing \\
   &  (hh:mm:ss)  &  (dd:mm:ss)  &     &  (s)  &      & (arcsec)  \\
\hline
Draco  &  17:20:19.651  &  +57:55:10.64   &   $y$  &  250 $\times$ 10  &  2014 Apr 27 &  1.26 \\
  &    &    &   $y$  &  250 $\times$ 7  &  2014 Apr 28 & 1.23 \\
  &    &    &   $y$  &  250 $\times$ 6  &  2015 Apr 21 & 0.86 \\
  &    &    &  $b$  &  500 $\times$ 5  &  2014 Apr 26 & 1.17 \\
  &    &    &  $b$  &  300 $\times$ 8  &  2014 Apr 28 & 1.12 \\
  &    &    &  $b$  &  500 $\times$ 5  &  2015 Apr 21 & 1.05 \\
  &    &    &  Ca  &  1800 $\times$ 5  &  2014 Apr 26 & 1.12 \\
  &    &    &  Ca  &  1200 $\times$ 7  &  2015 Apr 21 & 1.03 \\
  &    &    &  Ca  &  1200 $\times$ 5  &  2015 Apr 22 & 1.22 \\
Sextans  &  10:13:02.900  &  -01:36:53.00  &   $y$  &  180 $\times$ 5  &  2012 Dec 16 & 1.23 \\
  &    &    &   $y$  &  250 $\times$ 5  &  2015 Apr 21 & 0.86 \\
  &    &    &  $b$  &  360 $\times$ 5  &  2012 Dec 16 & 1.23 \\
  &    &    &  $b$  &  500 $\times$ 5  &  2015 Apr 21 & 0.86 \\
  &    &    &  Ca  &  1800 $\times$ 5  &  2012 Dec 16 & 1.41 \\
  &    &    &  Ca  &  1200 $\times$ 4  &  2015 Apr 21 & 1.19 \\
  &    &    &  Ca  &  1200 $\times$ 3  &  2015 Apr 21 & 1.19 \\
  &    &    &  Ca  &  1200 $\times$ 5  &  2015 Apr 22 & 0.85 \\
Canes Venatici I &  13:28:08.102  &  +33:33:36.49   &   $y$  &  250 $\times$ 5  &  2015 Apr 22 & 1.20 \\
 (CVnI)   &    &    &  $b$  &  500 $\times$ 5  &  2015 Apr 22 & 1.17 \\
  &    &    &  Ca  &  1200 $\times$ 10  &  2015 Apr 22 & 1.04 \\
  \hline
\end{tabular}
\end{center}
\label{t:log}
\end{table*}

In a $\Lambda$ cold dark matter universe, galaxies formed as part of the local overdensity in the matter distribution \citep[e.g.,][]{blu82,pee82}.
Galaxies are created via the agglomeration of numerous small protogalaxies as basic building blocks \citep[e.g.,][]{sea78,fon11}.
Many of such protogalaxies can escape from takeover by bigger galaxies and then develop independently into present-day dwarf galaxies. 
Hence, dwarf satellites around large galaxies are living fossils, containing key information on how galaxies form and evolve \citep[e.g.,][]{ric05}.

Spectroscopic observations of dwarf spheroidal galaxies (dSphs) revealed sizable metallicity spreads among their constituent stars \citep{kir08,kir11a,nor08}.
The observed internal abundance inhomogeneity in dSphs can be produced by the following two mechanisms: 
($a$) If the dSphs were able to keep their gas against their earliest supernova explosions, the gas could turn into stars with higher metallicity than the first generations. 
In this case, the newer stars are expected to be centrally concentrated within galaxies, leading to a rather strong radial gradient in metallicity. 
($b$) If dSphs were made by the agglomeration of even smaller progenitors (i.e., protodwarfs), the merger-induced star formation and accompanying chemical enrichment can cause abundance inhomogeneity. 
In this case, the radial gradient in metallicity is expected to be weak.
For a better understanding of the chemical enrichment history in dSphs, examining the structural feature of stellar populations as a function of metallicity is needed.

This notion calls for an examination on the dSph structural complexity as a function of metallicity, which we refer to as a ``chemostructural study.'' 
Obviously, the main obstacle to such a chemostructural study is the lack of stars with spectroscopic metallicities, and the presence of the possible chemostructural pattern across dSphs could be severely hampered by the small number statistics. 
In order to overcome this drawback, we use the Ca filter, designed to measure Ca $_{\rm II}$ H and K lines, as an alternative way to obtain stellar metallicity for a much larger sample than spectroscopy \citep[e.g.,][]{att91,att95,twa91,bai96,rey00,jlee09,cal07,cal11,lee13}.
Recent studies of Milky Way globular clusters (MWGCs) show that the Ca filter in the $hk$ index [$\equiv($Ca$-b)-(b-y)$] is very sensitive to the abundance of calcium \citep{han15,lim15}. 
Hence, the $hk$ index can be a powerful photometric metallicity discriminator. 
It is important to note that calcium can be supplied mostly by SNe II \citep{tim95}, in contrast to the lighter elements that can be increased by dredge-up and/or pollution from the intermediate-mass asymptotic-giant-branch stars. 
Therefore, the $hk$ index enables us to investigate primordial chemical inhomogeneity avoiding the influences by subsequent galactic evolution that produces the lighter elements.

To expand the new cost-effective technique, already confirmed for MWGCs, to Galactic companions, we obtained Subaru/Suprime-Cam Ca$-by$ photometry for Draco, Sextans, and Canes Venatici I (CVnI). 
The Draco dSph was first discovered on the Palomar Observatory Sky Survey in 1955 \citep{wil55}.
It is located 76 kpc away from the Galactic center and its stellar mass is 2.9 $\times$ 10$^5$\,$M$$_{\odot}$ \citep{mcc12}. 
In contrast to the early-discovered Draco, Sextans was recently found and classified into classical dSphs \citep{irw90}. 
It is located 86 kpc away from the Galactic center, occupying a large area  on the sky ($r$$_{h}=$ 695 pc) with very low surface brightness (${\mu}_V$\,=\,27.1 ${\rm mag}/{\square}\arcsec$) similar to that of ultra-faint dwarf galaxies \citep[UFDs;][]{mcc12}.
Its stellar mass is 4.4 $\times$ 10$^5$\,$M$$_{\odot}$. 
CVnI was found as stellar overdensity from SDSS data release 5 \citep{zuc06} and classified as one of the brightest UFDs. 
It lies at a distance of 218 kpc from the Galactic center and its stellar mass is 2.3 $\times$ 10$^5$\,$M$$_{\odot}$ \citep{mcc12}.

All these dSphs harbor relatively metal-poor, old stellar populations as evidenced by their color$-$magnitude diagrams (CMDs) such as the steep red-giant-branch (RGB) slope, the well-developed horizontal-branches (HBs), and the magnitude difference between HB and main-sequence turn-off. 
The rich population of RR Lyrae variables also indicates the presence of old stellar populations since the RR Lyrae stars are good tracers for metal-poor, old stellar populations. 
Although the majority of stars are metal-poor, the wide spread in abundance is reported from spectroscopic studies \citep[e.g.][]{kir10}. 
The inhomogeneity in abundance holds for even heavy elements such as Fe, Ca, which can be products of supernova explosions. 
Hence, the study of ``chemostructural'' property for these dSphs will give us clues to understanding the early phase of the evolution for the galaxies.

In this series of papers, we will explore the chemostructural features and stellar populations in dSphs using a large sample of stars with known metallicities obtained from our Ca$-by$ photometry. 
The galaxies of interest include Draco, Sextans, CVnI, Sculptor, and Fornax. 
This is the first paper to reassure an advantage of the Ca filter and the $hk$ index and to present the structural features as a function of stellar metallicity in Draco, Sextans, and CVnI. 
This paper is organized as follows. 
In Section 2, we present details on the observation, data reduction, and the characteristics of the Ca filter. 
Section 3 describes the CMDs and data analysis. 
We divide RGB stars into two groups, metal-poor and metal-rich, based on the $hk$ index and investigate spatial distributions of two different metallicity groups in Section 4. 
The summary and discussion are given in Section 5.

\section{Observation and Data Reduction} \label{sec:data}

We have performed photometric observations with the Suprime-Cam on the Subaru telescope using the Ca, $b$, and $y$ filters during the three observing runs from 2012 December to 2015 April for three dwarf galaxies, Draco, Sextans, and CVnI. 
The Suprime-Cam provides a 34{\arcmin} $\times$ 27{\arcmin} field of view with a pixel scale of 0{\farcs}2 and consists of 10 (5$\times$2 array) 2k $\times$ 4k CCDs \citep{miy02}. 
The observational details are summarized in Table~\ref{t:log}.

\begin{figure}
        \centering
                \includegraphics[width=7.95cm,clip=false]{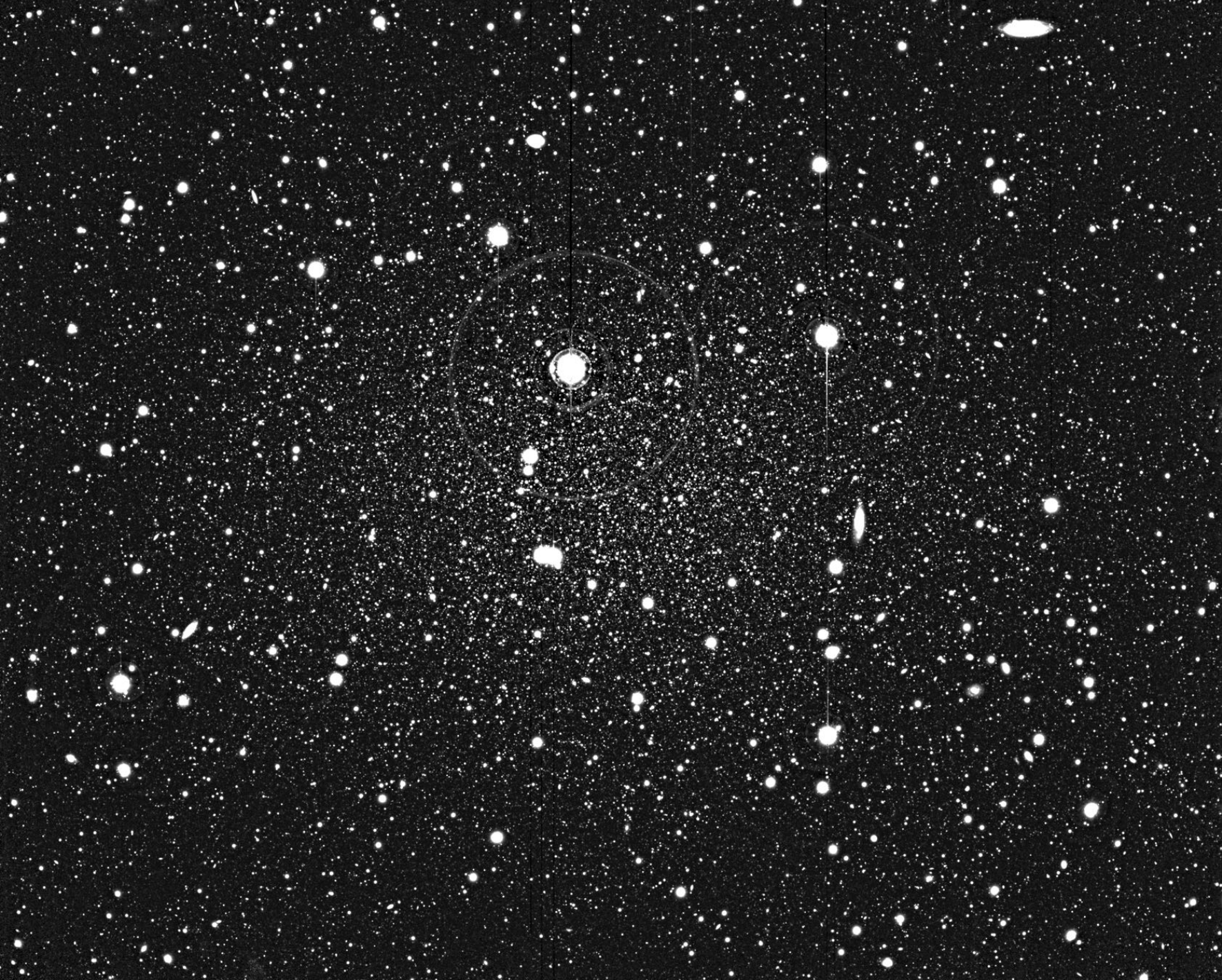}
                
                \includegraphics[width=7.95cm,clip=false]{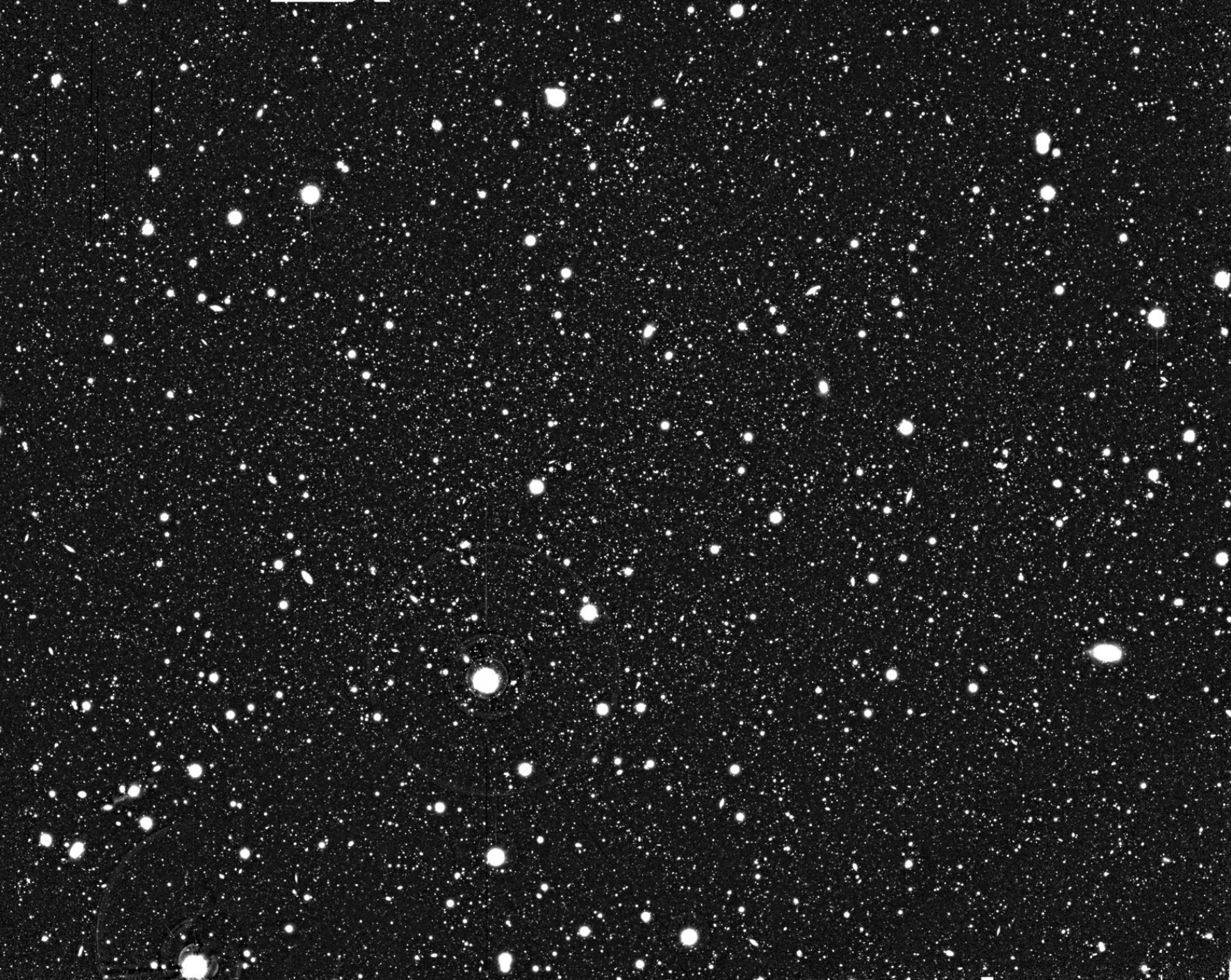}
                
                \includegraphics[width=7.95cm,clip=false]{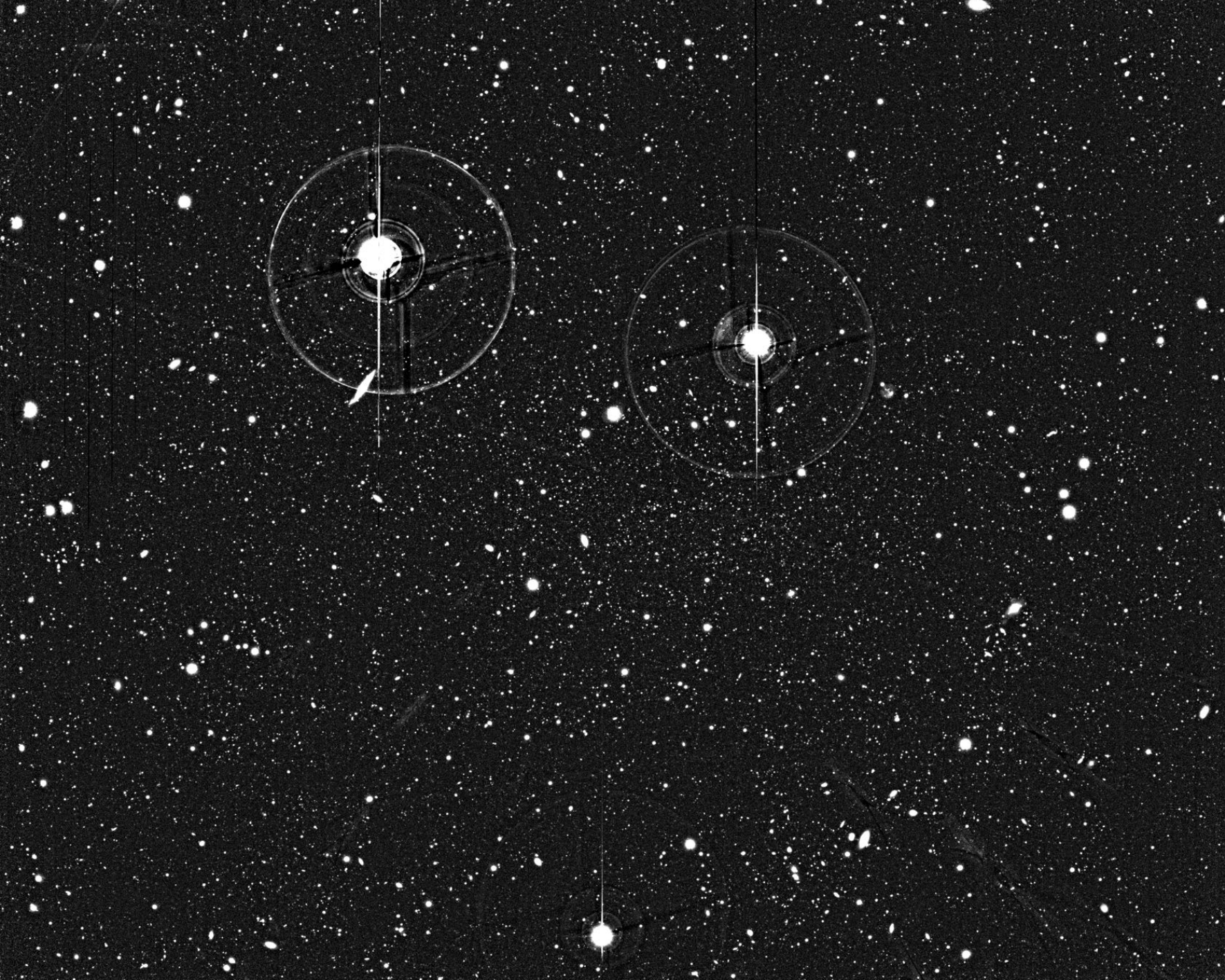}
        \caption{Combined $y$-band images of Draco (upper), Sextans (middle), and CVnI (bottom) dSphs, respectively, taken by Suprime-Cam. Each image covers a 34{\arcmin} $\times$ 27{\arcmin} field of view. \label{fig:f_images}}
\end{figure}

The raw data were preprocessed by SDFRED~1 and SDFRED~2, a semiautomated pipeline software for Suprime-Cam \citep{yag02,ouc04}. 
All images were bias-subtracted, trimmed, and flat-fielded in the same manner as single chip reduction. 
The master flat frames for each filter were obtained from the median combination of the object images. 
The distortion by a fast focus ({\it f}/2.0) was corrected with fourth-order polynomial fitting. 
The point-spread-function (PSF) size of all frames were equalized to have a common FWHM value through Gaussian smoothing before coadding. 
The interpolated sky counts of 12{\farcs}8 $\times$ 12{\farcs}8 meshes were used to subtract the sky background. 
While the frames were combining, the relative position and throughputs among frames were adjusted using the calculation from the common stars. 
We then obtained combined images for each filter.
Figure~\ref{fig:f_images} presents examples of the combined $y$-band images for the three dSphs.

\begin{figure}
        \centering
                \includegraphics[width=8.75cm,clip=false]{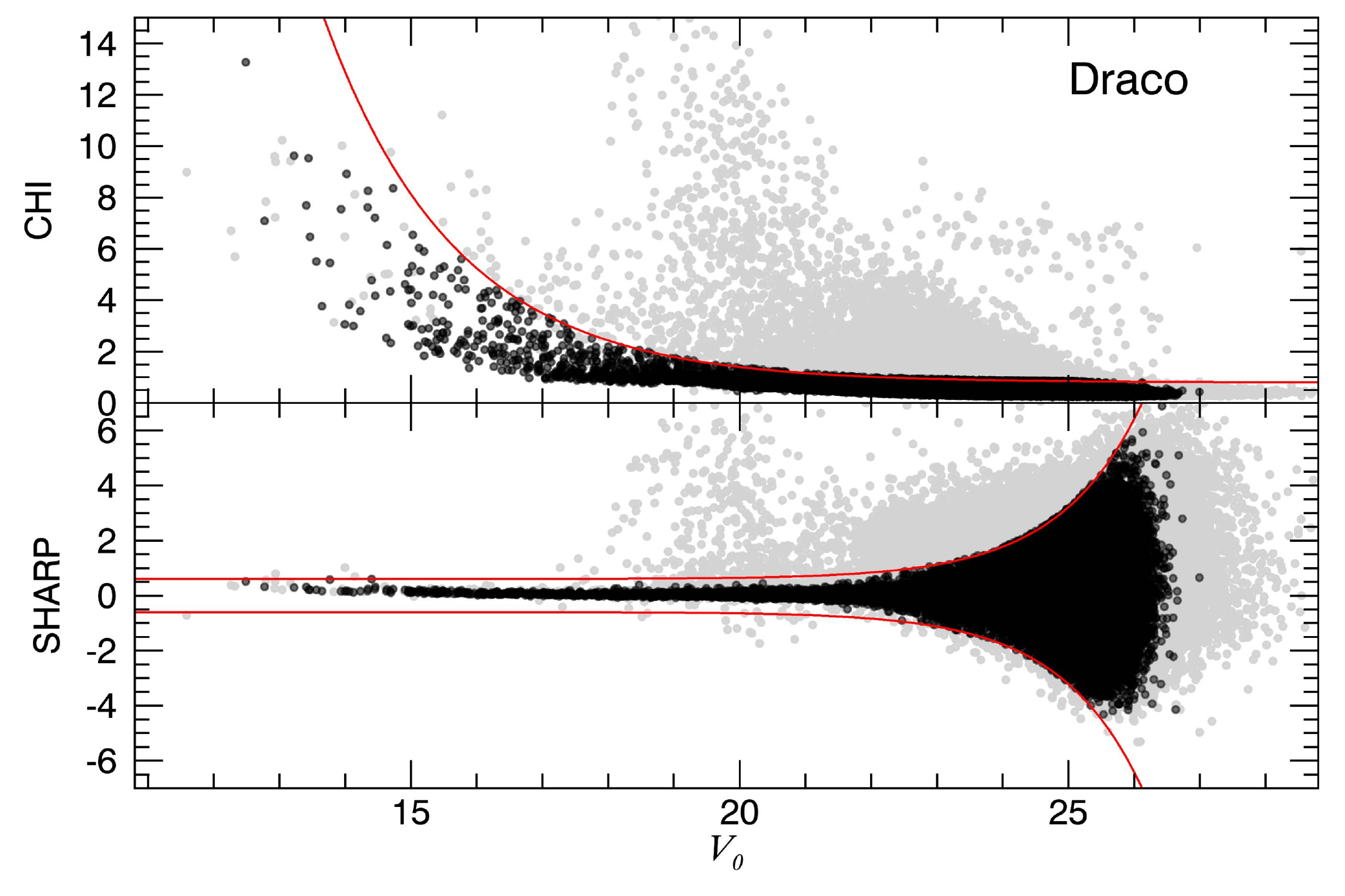}
                
                \includegraphics[width=8.75cm,clip=false]{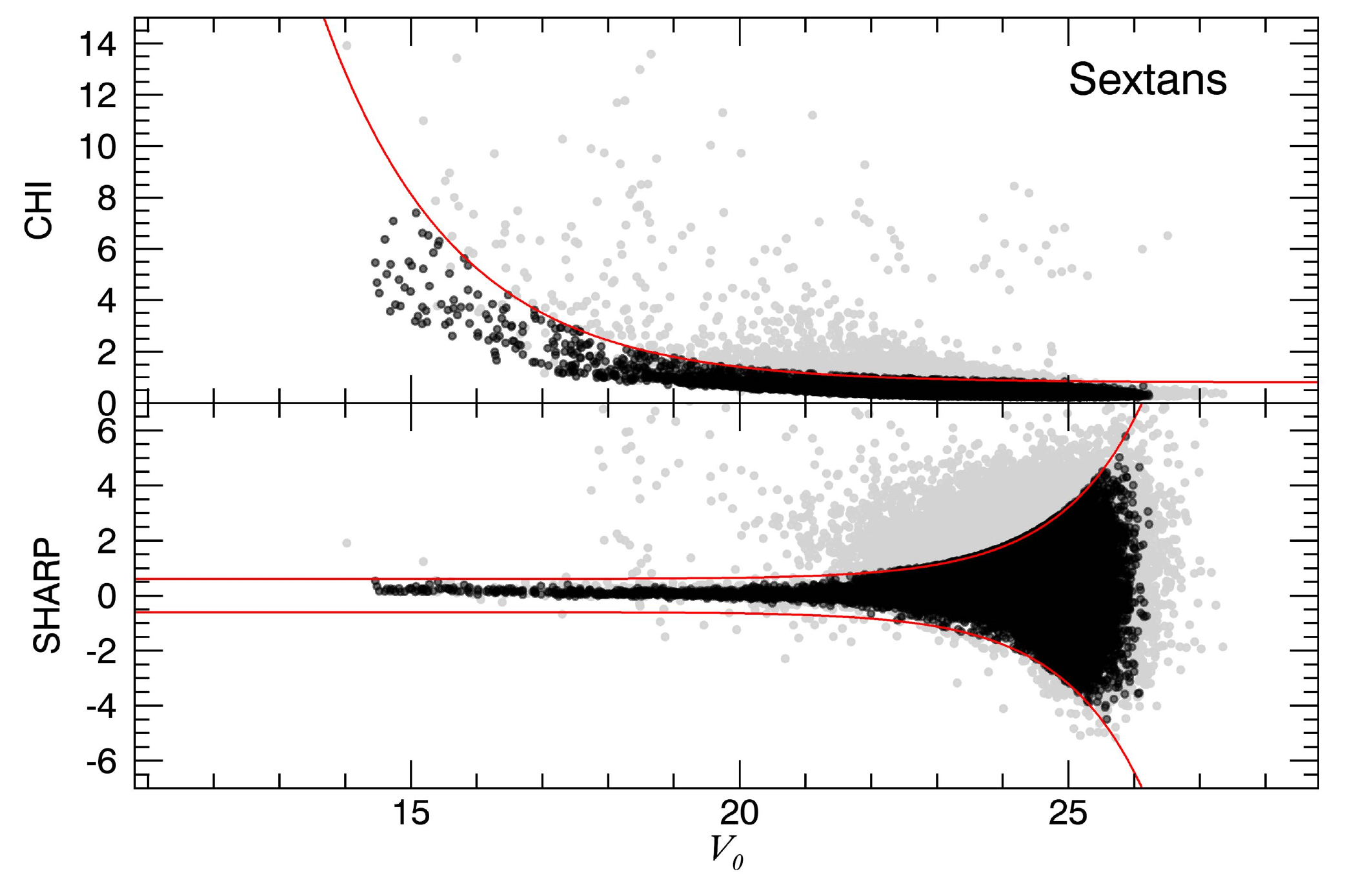}
                
                \includegraphics[width=8.75cm,clip=false]{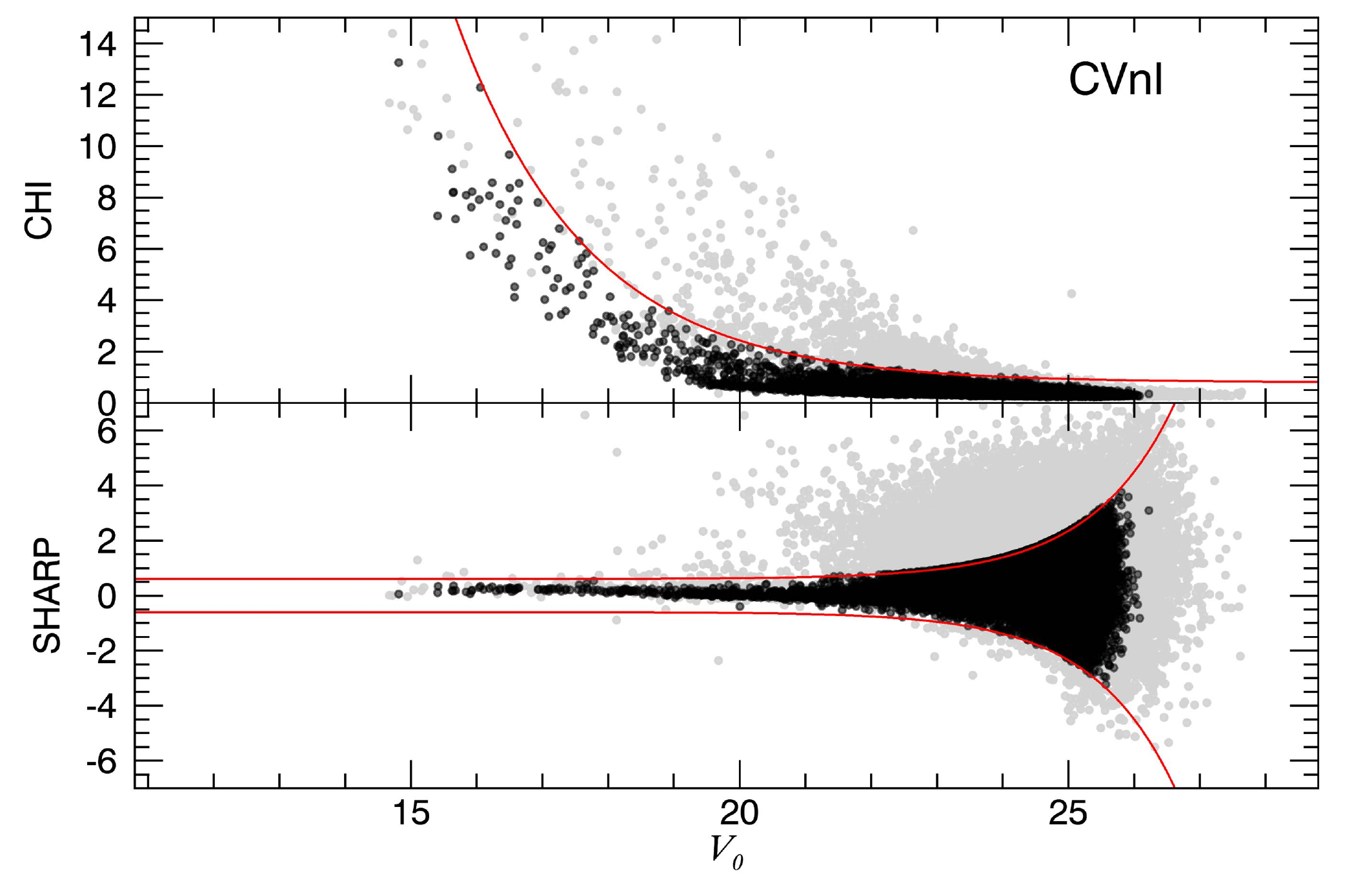}
        \caption{DAOPHOT parameters (CHI and SHARP) as a function of $V$ magnitude for the three dSphs. The red solid lines denote the nonstellar object rejection criteria. The black dots represent the selected stellar objects, while the gray dots represent the rejected nonstellar objects. \label{fig:f_cut}}
\end{figure}

We detected point sources from the combined images using DAOPHOT/ALLSTAR \citep{ste87} to make an input star list for ALLFRAME \citep{ste94}.
The PSF models were constructed for all individual images by DAOPHOT using the best-matched analytic approximation. 
Due to the large field of view, the PSF models vary quadratically in the frames. 
We performed PSF photometry for all individual images simultaneously using ALLFRAME.
Then, the point sources were matched up by DAOMATCH/DAOMASTER \citep{ste93} and the aperture corrections were applied using DAOGROW \citep{ste90}.

The instrumental Ca, $b$, and $y$ magnitudes were converted to the standard system using the MWGCs M5 and M15 because the standard stars are too bright for the Suprime-Cam. 
These GCs were observed with the du Pont 2.5 m telescope using the same new Ca-$by$ filter set and transformed to the standard Ca-$by$ photometric system. 
In order to calibrate the new Ca filter on the du Pont telescope to the standard Ca magnitude system \citep{att91}, not only Ca-$by$ standard stars \citep{att91,att98} but also open clusters such as NGC~3680, NGC~6253, and NGC~5822 were used \citep{twa03,att04,car11}.
The astrometric standard stars in the USNO B1 catalog were used to compute the astrometric solutions \citep{mon03}. 
The typical rms of the astrometric solutions is at the level of 0{\farcs}2.
Galactic extinction for each star in the three dSphs was corrected based on the dust map from \citet{sch11} and the extinction law provided by \citet{car89}.
To minimize nonstellar objects, we use the CHI and SHARP parameters from DAOPHOT (Figure~\ref{fig:f_cut}).
The magnitudes and errors for Ca, $b$, and $y$ filters are given in Table~\ref{t:data_drac} (Draco), Table~\ref{t:data_sext} (Sextans), and Table~\ref{t:data_cvni} (CVnI).

To improve the photometric quality and galaxy member star selection accuracy using multiband photometry, we reprocess the additional $V$ and $I$ images from the Subaru archive. 
The archival $V$- and $I$-band images for Draco (S. Okamoto 2020, in preparation), Sextans \citep{oka17}, and CVnI \citep{oka12}, taken between 2005 and 2008, were reanalyzed in the same manner as our Ca-$by$ reduction.
The $V$ and $I$ magnitudes were calibrated against the standard star catalog provided by \citet{ste00}. 
The number of stars presented in our study is less than that in other studies using the same data sets in \citet{oka12} for CVnI, \citet{oka17} for Sextans, and S. Okamoto et al. (2020, in preparation) for Draco because the $V$ and $I$ images overlapped with Ca photometry are reprocessed.

The original Ca filter was centered at 3995\,{\AA} with an FWHM of approximately 90\,{\AA} and designed to measure the Ca $_{\rm II}$ H and K lines \citep{att91}. 
Since the blue wing of the filter is close to the strong CN band at 3883\,{\AA}, we modified the Ca filter to avoid possible CN contamination and best measure the calcium abundances \citep{lee13}. 
The blue wing of the modified filter is slightly moved toward the longer wavelength and has a narrower FWHM (76\,\AA) with steeper wings in both blue and red compared to the original Ca filter. 
Therefore, the modified filter is practically measuring ionized calcium H and K lines traditionally used to calibrate the metallicity scale for MWGCs. 
This modified Ca filter has been used for our survey of MWGCs and dwarf galaxies. 
The $hk$ index is confirmed to detect a small difference in calcium abundance for RGB stars of MWGCs \citep{lee13,han15,lim15}, and this study applies for the first time the scheme to external dwarf galaxies.

\section{The Color$-$Magnitude Diagrams} \label{sec:cmds}

\begin{table*}
\footnotesize
\caption{Photometric Measurements for the Point Sources in Draco}
\begin{center}
\begin{tabular}{lcccccccrr}
\hline\hline

R.A.  &  Decl.  &  Ca  &  Ca Error  &  $b$ & $b$ Error & $y$ & $y$ Error & CHI~~ & SHARP \\

\hline

  260.489864 & 57.781853 & 14.183 & 0.004 & 12.762 & 0.012 & 12.368 & 0.004 & 8.987 & -0.716\\
  260.300760 & 57.825928 & 15.643 & 0.001 & 13.380 & 0.003 & 12.652 & 0.002 & 4.963 & 0.203\\
  259.858303 & 57.838948 & 14.626 & 0.002 & 13.134 & 0.006 & 12.685 & 0.006 & 13.266 & 0.511\\
  259.866306 & 57.774826 & 15.439 & 0.001 & 13.558 & 0.003 & 12.903 & 0.002 & 7.091 & 0.320\\
  260.031487 & 57.817089 & 14.274 & 0.003 & 13.286 & 0.006 & 12.977 & 0.002 & 10.229 & 0.211\\
  260.489159 & 58.050763 & 14.426 & 0.002 & 13.460 & 0.002 & 13.035 & 0.003 & 7.839 & 0.564\\
  259.970845 & 58.031147 & 14.366 & 0.003 & 13.413 & 0.003 & 13.054 & 0.002 & 5.695 & 0.342\\
  260.022305 & 57.862427 & 15.271 & 0.001 & 13.719 & 0.004 & 13.133 & 0.002 & 6.704 & 0.381\\
  260.572728 & 57.816874 & 14.722 & 0.001 & 13.644 & 0.004 & 13.152 & 0.002 & 9.602 & 0.263\\
  260.117631 & 57.885508 & 15.518 & 0.032 & 12.556 & 0.014 & 13.166 & 0.032 & 10.043 & 1.677\\

\hline
\end{tabular}
\end{center}
\label{t:data_drac}
\tablecomments{This table is available in its entirety in a machine-readable form.}
\end{table*}

\begin{table*}
\footnotesize
\caption{Photometric Measurements for the Point Sources in Sextans}
\begin{center}
\begin{tabular}{lcccccccrr}
\hline\hline

R.A.  &  Decl.  &  Ca  &  Ca Error  &  $b$ & $b$ Error & $y$ & $y$ Error & CHI~~ & SHARP \\

\hline

  153.413563 & -1.701420 & 15.437 & 0.001 & 14.094 & 0.053 & 13.656 & 0.015 & 13.914 & 1.908\\
  153.185806 & -1.726351 & 16.926 & 0.001 & 15.249 & 0.001 & 14.450 & 0.002 & 4.686 & 0.335\\
  153.419309 & -1.433601 & 16.283 & 0.001 & 14.927 & 0.001 & 14.473 & 0.003 & 5.463 & 0.532\\
  153.412262 & -1.565279 & 15.835 & 0.001 & 14.877 & 0.005 & 14.564 & 0.002 & 4.279 & 0.164\\
  153.038128 & -1.663527 & 16.087 & 0.001 & 15.052 & 0.002 & 14.635 & 0.002 & 6.370 & 0.156\\
  153.175699 & -1.794828 & 16.074 & 0.001 & 15.033 & 0.001 & 14.675 & 0.002 & 5.020 & 0.238\\
  153.523789 & -1.569433 & 16.715 & 0.001 & 15.240 & 0.001 & 14.694 & 0.002 & 3.570 & 0.169\\
  153.470283 & -1.610698 & 17.327 & 0.001 & 15.322 & 0.002 & 14.703 & 0.003 & 7.084 & 0.223\\
  153.155947 & -1.606000 & 16.587 & 0.001 & 15.216 & 0.001 & 14.706 & 0.002 & 5.396 & 0.242\\
  153.197029 & -1.638868 & 16.401 & 0.001 & 15.209 & 0.002 & 14.795 & 0.002 & 3.829 & 0.142\\

\hline
\end{tabular}
\end{center}
\label{t:data_sext}
\tablecomments{This table is available in its entirety in a machine-readable form.}
\end{table*}

\begin{table*}
\footnotesize
\caption{Photometric Measurements for the Point Sources in CVnI}
\begin{center}
\begin{tabular}{lcccccccrr}
\hline\hline

R.A.  &  Decl.  &  Ca  &  Ca Error  &  $b$ & $b$ Error & $y$ & $y$ Error & CHI~~ & SHARP \\

\hline

  202.153692 & 33.686321 & 16.828 & 0.001 & 15.350 & 0.003 & 14.807 & 0.008 & 8.148 & 0.682\\
  202.263594 & 33.740072 & 16.611 & 0.001 & 15.394 & 0.002 & 14.892 & 0.004 & 11.680 & 0.011\\
  202.079124 & 33.757789 & 16.295 & 0.001 & 15.360 & 0.004 & 14.927 & 0.009 & 13.254 & 0.056\\
  202.219941 & 33.481131 & 16.366 & 0.001 & 15.289 & 0.007 & 14.945 & 0.005 & 14.390 & 0.001\\
  201.910729 & 33.565612 & 16.769 & 0.001 & 15.373 & 0.005 & 14.965 & 0.008 & 9.991 & 0.356\\
  201.669906 & 33.543535 & 16.415 & 0.002 & 15.506 & 0.005 & 15.133 & 0.008 & 11.584 & 0.527\\
  202.329525 & 33.686368 & 18.314 & 0.001 & 16.077 & 0.001 & 15.299 & 0.002 & 10.646 & -0.039\\
  201.881973 & 33.715510 & 16.557 & 0.001 & 15.682 & 0.001 & 15.312 & 0.002 & 11.428 & 0.205\\
  201.964234 & 33.492423 & 17.361 & 0.001 & 15.875 & 0.001 & 15.350 & 0.001 & 7.116 & 0.084\\
  201.837374 & 33.692573 & 18.055 & 0.001 & 15.943 & 0.002 & 15.371 & 0.002 & 13.208 & 0.182\\

\hline
\end{tabular}
\end{center}
\label{t:data_cvni}
\tablecomments{This table is available in its entirety in a machine-readable form.}
\end{table*}

\begin{figure*}[ht!]
\epsscale{1.05}
\plotone{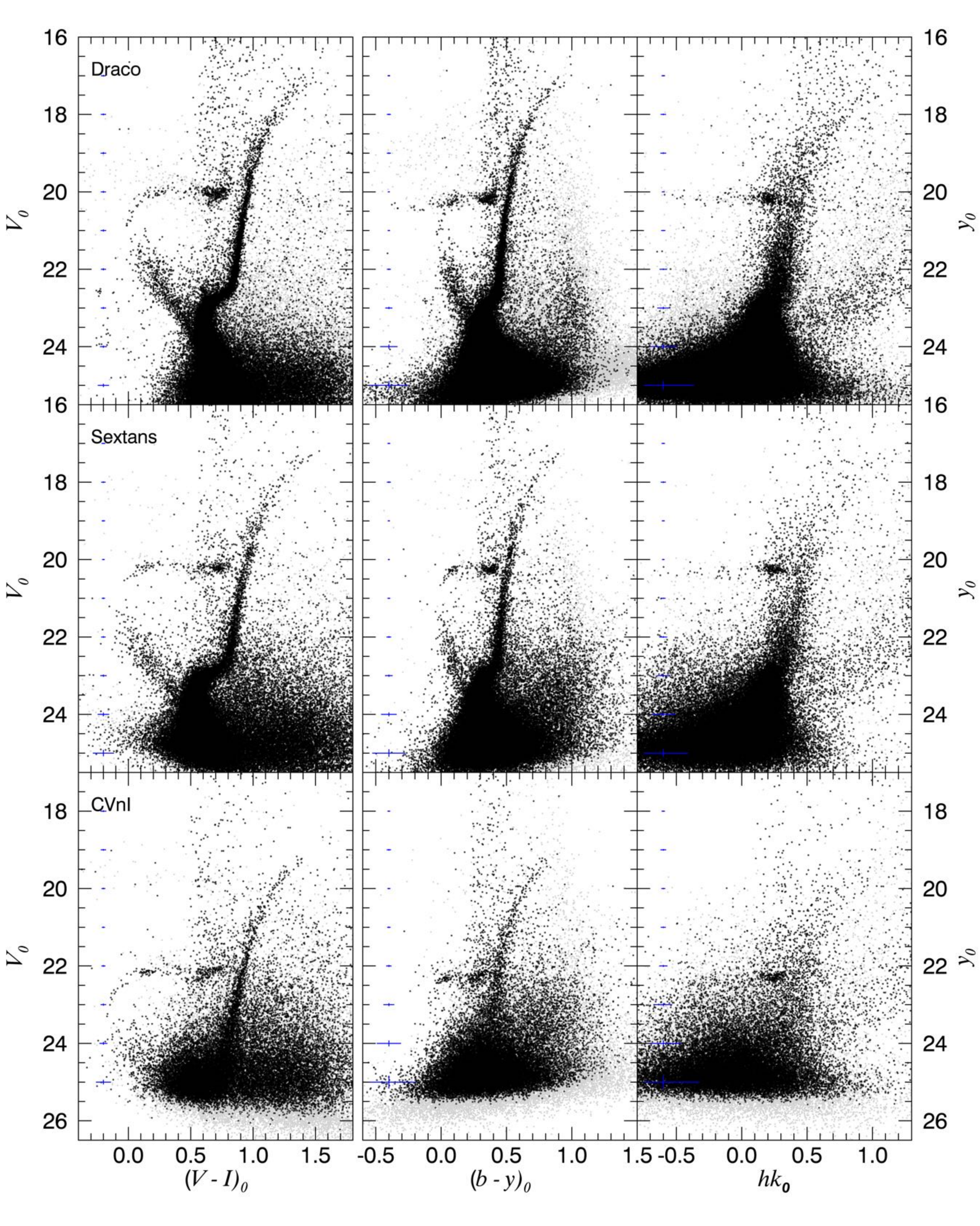}  
\caption{The CMDs for Draco (top), Sextans (middle), and CVnI (bottom) in $(V, V-I)$, $(y, b-y)$, and $(y, hk)$ planes. The error bars denote the photometric errors at given magnitude bins. The nonstellar objects rejected by CHI and SHARP are shown as gray dots.\\ \label{fig:f_cmd_all}}
\end{figure*}

\begin{figure*}[ht!]
\epsscale{1.05}
\plotone{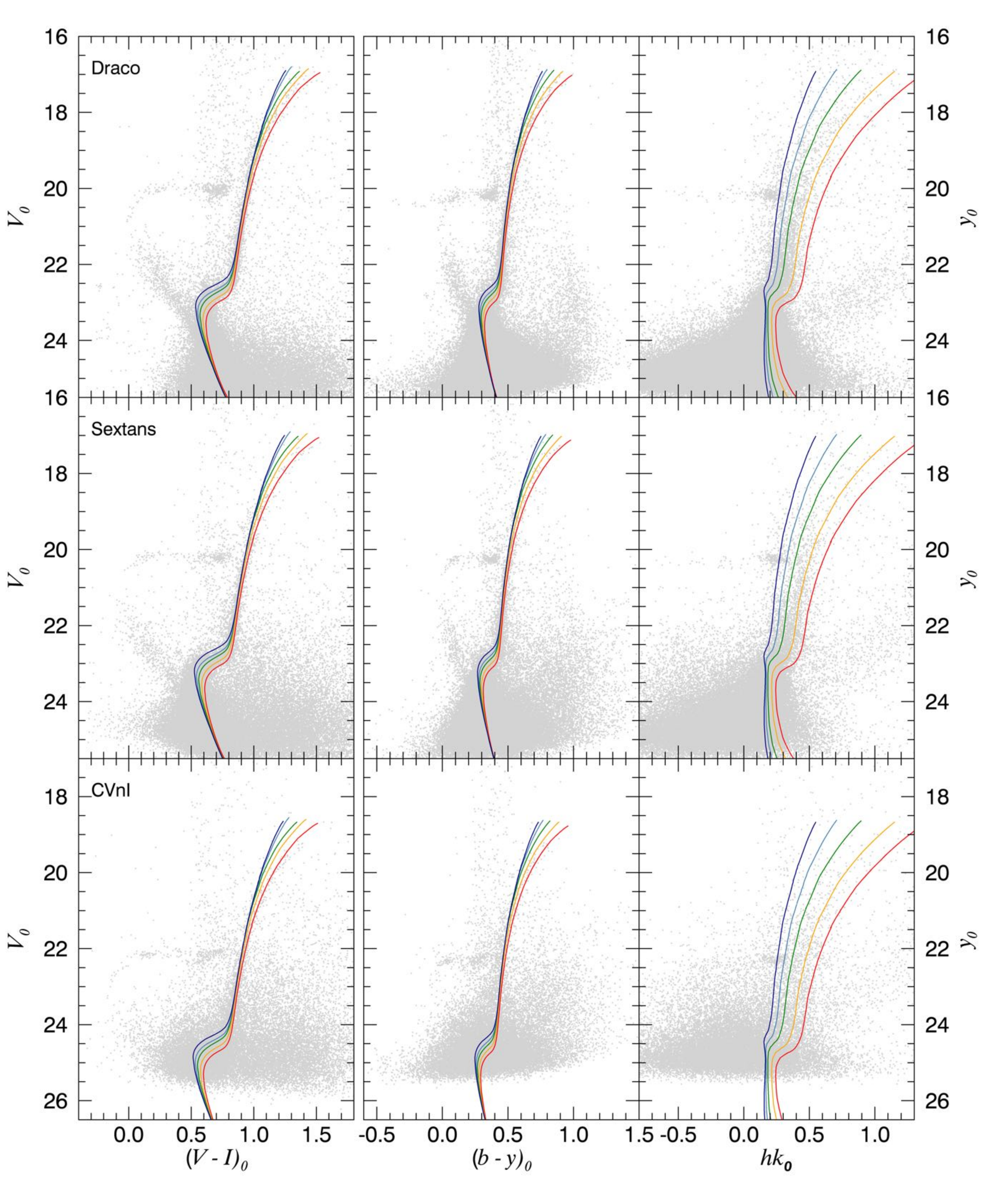}  
\caption{The $Y^{2}$ isochrones overlaid on the CMDs for the three dSphs. The metallicity of isochrones are $-$3.25 (navy), $-$2.75 (blue), $-$2.25 (green), $-$1.75 (orange), and $-$1.50 (red). The different $\alpha$ values are considered ([$\alpha$/Fe]\,=\,0.4 for [Fe/H]\,=\,$-$3.25 and $-$2.75, [$\alpha$/Fe]\,=\,0.2 for [Fe/H]\,=\,$-$2.25, and  [$\alpha$/Fe]\,=\,0.0 for [Fe/H]\,=\,$-$1.75 and $-$1.50). The all isochrones are for 12.5 Gyr.\\  \label{fig:f_cmd_iso}}
\end{figure*}

Figure~\ref{fig:f_cmd_all} shows the CMDs of Draco (upper), Sextans (middle), and CVnI (bottom). 
From the left column to the right, the $V$ versus $V-I$, $y$ versus $b-y$, and $y$ versus $hk$ CMDs are presented. 
The gray dots are possible nonstellar objects selected by DAOPHOT parameters (CHI and SHARP).
In order to quantify the degree of the metallicity spread among RGB stars in the dSphs, we overlaid the $Y^{2}$ isochrones \citep[version 3;][]{yi08} on the observed CMDs in Figure~\ref{fig:f_cmd_iso}. 
We use the isochrones of 12.5\,Gyr with [Fe/H]\,=\,$-$3.25, $-$2.75, $-$2.25, $-$1.75, and $-$1.50. 
The $\alpha$-element enhancement is applied following the abundance trend found in the three dSphs \citep{kir11a}: [$\alpha$/Fe]\,=\,0.4 for [Fe/H]\,=\,$-$3.25 and $-$2.75; [$\alpha$/Fe]\,=\,0.2 for [Fe/H]\,=\,$-$2.25; and [$\alpha$/Fe]\,=\,0.0 for [Fe/H]\,=\,$-$1.75 and $-$1.50. 
We do not apply the younger age isochrones for the metal-richer populations but instead single-age isochrones are used. 
This is because the star formation periods of these dSphs are relatively narrow at old ages \citep[$\simeq$ 10$-$13 Gyr;][]{lee09,kir11a,wei14} and the effect of this short age interval ($\leq$ 3 Gyr) on the $hk$ index of RGB sequence is negligible\footnote{The examination using the isochrones for the $hk$ index shows that the effect of 0.1 dex difference in metallicity on the RGB is comparable to the effect of a 6 Gyr difference in age}.
Since the $Y^{2}$ isochrones do not provide the Ca$-by$ system colors, we calculate the $hk$ index and the $b-y$ color from the color$-$temperature relations given by Castelli\footnote{http://wwwuser.oats.inaf.it/castelli/colors.html} based on ATLAS9 models \citep{cas06}.

\begin{figure*}[ht!]
\epsscale{1.05}
\plotone{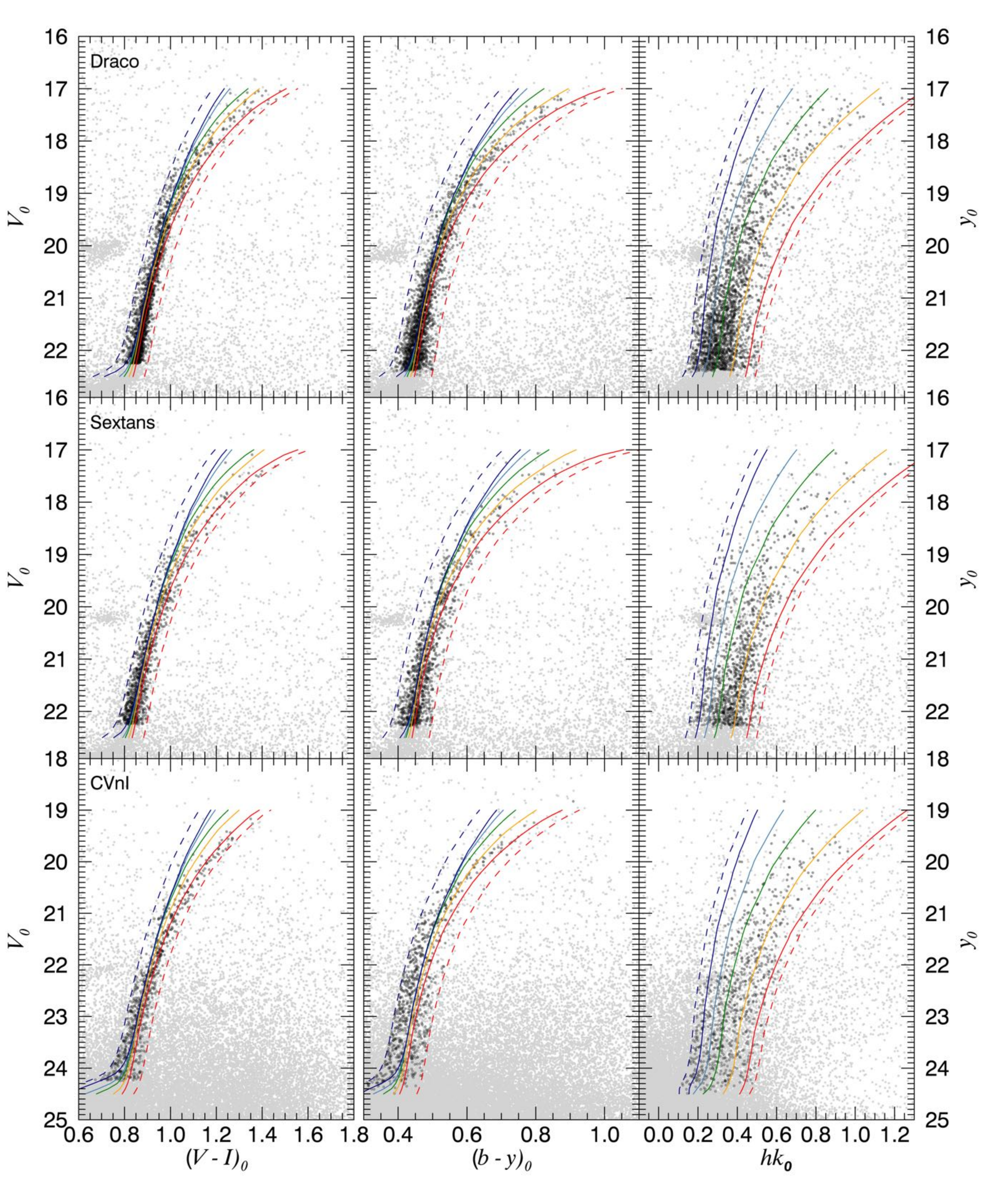}  
\caption{RGB member stars for the three dSphs. The solid lines are the same as the isochrones shown in Figure~\ref{fig:f_cmd_iso} and the dashed lines are 0.05 mag bluer than [Fe/H]\,=\,$-$3.25 isochrones and 0.05 mag redder than [Fe/H]\,=\,$-$1.50 isochrones. The black dots are the most probable RGB member stars located between two dashed lines for all of the three colors. \\  \label{fig:f_rgb_mem}}
\end{figure*}

\begin{figure*}[ht!]
\epsscale{1.175}
\plotone{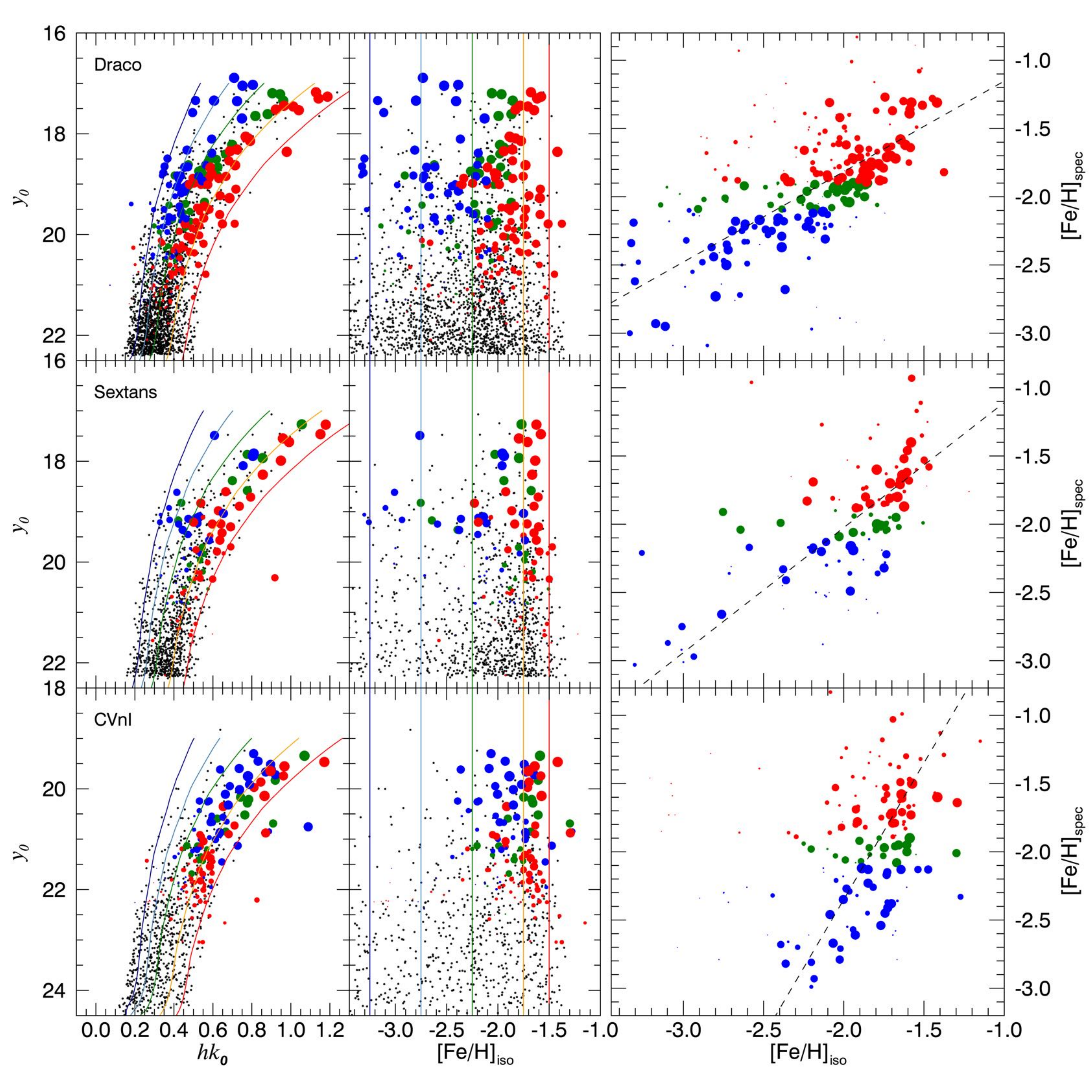}  
\caption{Comparison of our Ca$-by$ photometry with the spectroscopic data from \citet{kir10}. The blue, green, and red dots are, respectively, for metal-poor ([Fe/H]$_{ \rm spec}$ $<$ $-$2.1), metal-intermediate ($-$2.1 $<$ [Fe/H]$_{ \rm spec}$ $<$ $-$1.9), and metal-rich ([Fe/H]$_{ \rm spec}$ $>$ $-$1.9) RGB stars. Left panels: The same as the $hk$ CMDs in Figure~\ref{fig:f_rgb_mem}. The navy, blue, green, orange, and red lines represent [Fe/H]\,=\,$-$3.25, $-$2.75, $-$2.25, $-$1.75, and $-$1.50 isochrones, respectively. Middle panels: the verticalized diagrams (see the text). Right panels: correlation between [Fe/H]$_{\rm iso}$ and the spectroscopic metallicity ([Fe/H]$_{ \rm spec}$). The size of the symbol is inversely proportional to the observational uncertainty. The black dashed lines indicate the  linear  orthogonal  regression  fit.\\ \label{fig:f_rgb_kirby}}
\end{figure*}

\begin{figure*}[ht!]
\epsscale{1.175}
\plotone{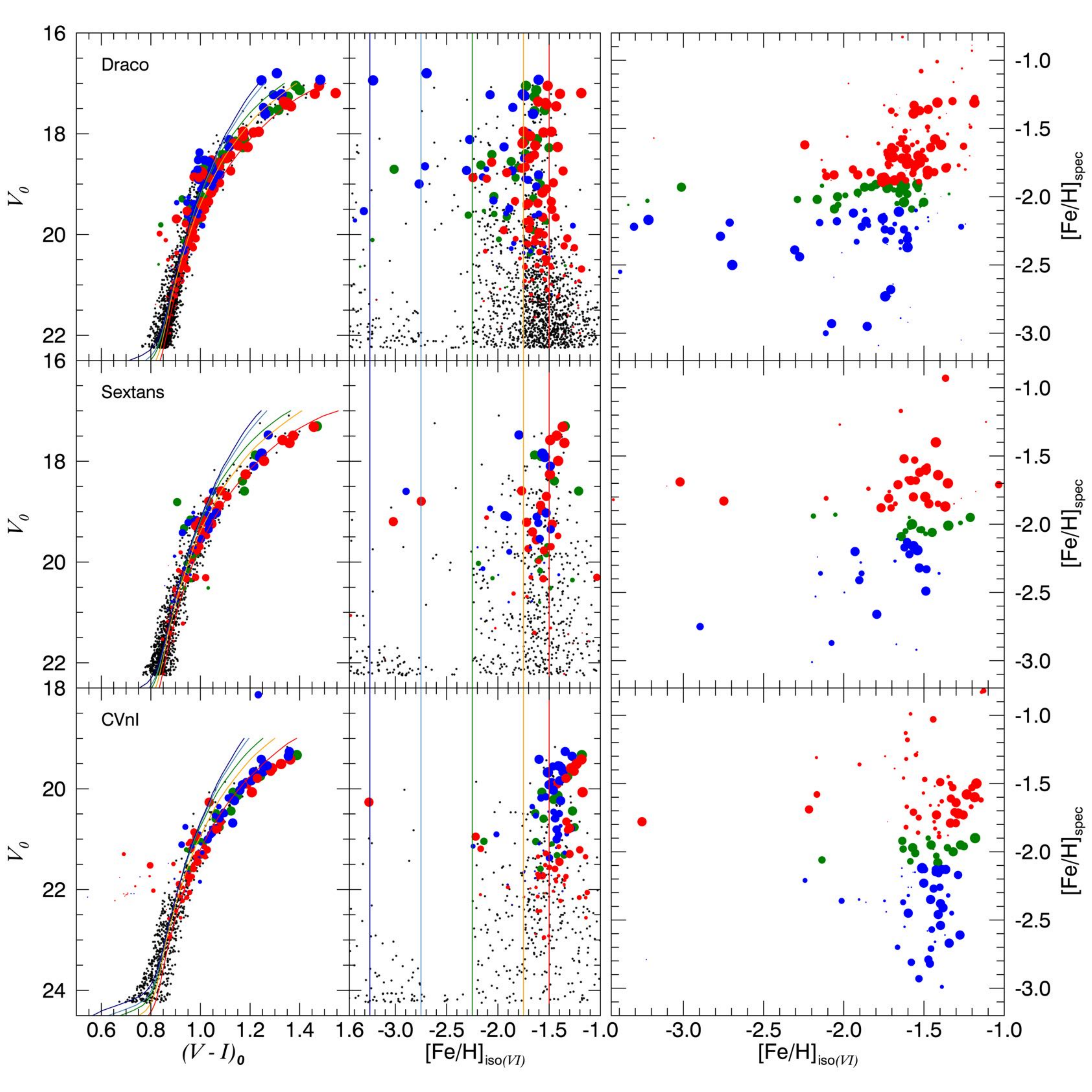}  
\caption{Same as Figure~\ref{fig:f_rgb_kirby}, but for $V$ and $I$ photometry.\\ \label{fig:f_rgb_kirby_vi}}
\end{figure*}

As shown in Figure~\ref{fig:f_cmd_all}, the Galactic field stars coincide with the RGB sequence of the galaxies for $V-I$ and $b-y$ colors.
For a better selection of member RGB stars and minimization of Galactic field star contamination, we take advantage of the Ca filter characteristics: the field stars that are usually more metal-rich have much larger $hk$ indices than the galaxies' member RGB stars. 
This is because the $hk$ index is more sensitive to the metallicity than the other two colors.
Using such property of the $hk$ index, the member RGB stars are determined in the following process.
We first match the isochrones on the observed CMDs considering the overall shape from the main sequence to the RGB for $V-I$, $b-y$, and $hk$ (Figure~\ref{fig:f_cmd_iso}). 
Then, we define the blue (red) boundary as 0.05 mag bluer (redder) than the most metal-poor (metal-rich) isochrone (Figure~\ref{fig:f_rgb_mem}). 
The quantity, 0.05 mag in $V-I$, $b-y$, and $hk$, is three times the typical photometric error for RGB stars at the HB level.
The stars enclosed by the blue and red boundaries of $V-I$, $b-y$, and $hk$ {\it in common} are selected as member RGB stars of the galaxies.

To verify the feasibility of the $hk$ index as a metallicity indicator, we compare our Ca$-by$ photometry and the spectroscopic metallicity data in the literature \citep{kir10}.
In the left panels of Figure~\ref{fig:f_rgb_kirby}, the bigger (smaller) dots denote the spectroscopic samples with small (large) errors.
The $hk$ index of a star gets redder as its metallicity gets higher at a given magnitude.
We define [Fe/H]$_{\rm iso}$ as the metallicity through interpolation (or extrapolation if needed) between two enveloping isochrones at the corresponding $y$ magnitude. 
In the middle panels, the RGB slope is verticalized by using [Fe/H]$_{\rm iso}$. 
In the right panels, [Fe/H]$_{\rm iso}$ positively correlates with spectroscopic [Fe/H] ([Fe/H]$_{\rm spec}$; \citealt{kir10}).
The strong correlation ensures that [Fe/H]$_{\rm iso}$ can be used as a photometric metallicity indicator.

\begin{figure*}[ht!]
\epsscale{1.175}
\plotone{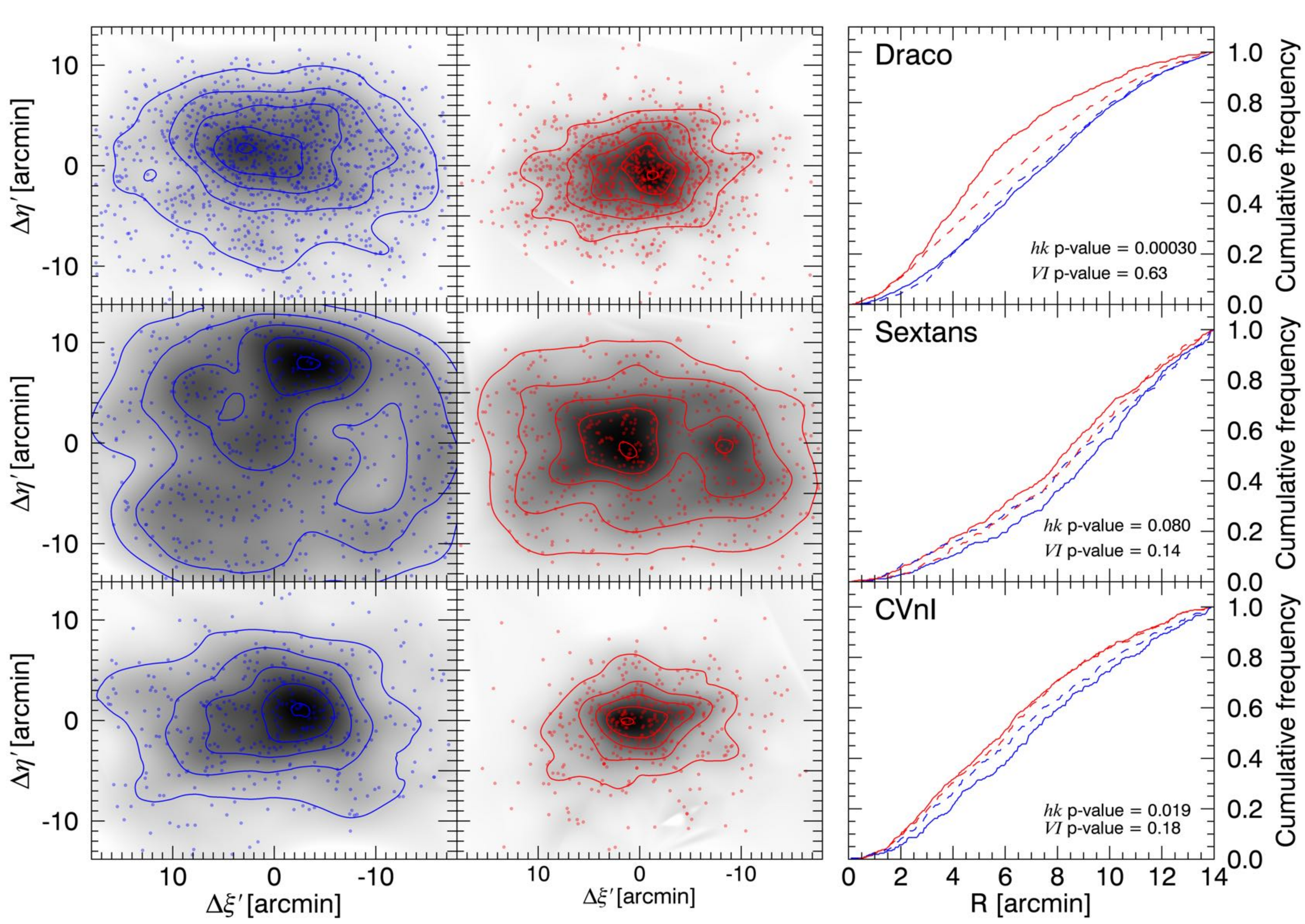}  
\caption{Spatial distribution of the metal-poor (left) and the metal-rich (center) RGB stars and the radial cumulative curves (right) for Draco (top), Sextans (middle), and CVnI (bottom). Left and middle panels: the RGB stars are divided into two groups, metal-poor (left; blue dots) and metal-rich (center; red dots). Each group has a similar numbers of stars. The background density plots and the overlaid contour plots are created by applying kernel density estimation (see the text).
Right panels: the solid blue and red lines are, respectively, for the bluer and redder RGB stars defined by the $hk$ index. The dashed blue and red lines are, respectively, for the bluer and redder RGB stars defined by the $V-I$ color. 
The cumulative distributions show that the redder RGB stars are more concentrated centrally than the bluer RGB stars. The redder RGB stars in the $V-I$ color (dashed lines) are weaker than those in the $hk$ index (solid lines). The $p$-value in the right panels denotes the KS probability, giving a significance of being equal.\\
 \label{fig:f_con_kde}} 
\end{figure*}

The origin of the differences in slopes in the [Fe/H]$_{\rm iso}$ versus [Fe/H]$_{\rm spec}$ plots of the three dSphs is not clear yet and to be investigated further with better calibrations between the photometric and spectroscopic metallicity measures (see Section~\ref{sec:summarydiscussion}). 
As for the photometric accuracy, aside from the observational uncertainties, some intrinsic characteristics may contribute to the dispersion of photometric metallicity, [Fe/H]$_{\rm iso}$.
The chromospheric emission in the Ca II K line makes the $hk$ index smaller \citep{att98}. 
Also, the Bond--Neff opacity affected stars show broad absorption at $\sim$4000\,{\AA}, making the $hk$ index greater \citep{bon69,att95,cal11}.

In Figure~\ref{fig:f_rgb_kirby_vi}, we plot the CMDs (left panels) and the verticalized CMDs (middle panels) for $V-I$ using the same method for the $hk$ index to check whether the metallicity derived from the $V-I$ isochrones ([Fe/H]$_{\rm iso(\it VI)}$) can be used to discriminate between metal-poor and metal-rich RGB stars. 
In the right panels, the correlation between [Fe/H]$_{\rm iso(\it VI)}$ and the spectroscopic metallicity is too weak. 
Hence, the two RGB groups divided in the [Fe/H]$_{\rm iso(\it VI)}$ are most likely cross-contaminated with each other and cannot represent the different metallicity groups.
The comparison of metallicity-discriminating powers between $hk$ and $V-I$ will be further discussed in Section \ref{sec:tworgb}.

\section{The chemostructures of RGB stars revealed by the $\MakeLowercase{hk}$ index} \label{sec:tworgb}

\subsection{Spatial Distribution of the Metal-poor and the Metal-rich RGB Stars}

In Figure~\ref{fig:f_con_kde}, we compare the metal-poor and the metal-rich RGB stars in Draco (upper), Sextans (middle), and CVnI (bottom) using the contours (left and center) and the cumulative curves (right). 
The RGB stars are divided into the two groups, the metal-poor and the metal-rich, based on the
[Fe/H]$_{\rm iso}$ to contain similar numbers of stars. Each group in Draco, Sextans, and CVnI contains approximately 1050, 520, and 380 stars, respectively. 
The left and center panels of Figure~\ref{fig:f_con_kde} display the spatial distributions of the metal-poor (blue) and the metal-rich (red) stars, respectively.
The right panels show the cumulative distributions of the metal-poor (blue solid line) and the metal-rich (red solid line) stars are shown. 
The dashed lines indicate the blue and the red RGB stars divided based on [Fe/H]$_{\rm iso(\it VI)}$ instead of [Fe/H]$_{\rm iso}$.

The structural parameters such as the center coordinate, half-light radius, position angle, and ellipticity are obtained from \citet{mcc12}. 
The equatorial spherical coordinates ($\alpha$ and $\delta$) of all objects in the observed fields are transformed to the rectangular coordinates ($\xi$ and $\eta$). 
In order to make its major axis to be parallel to the x-axis ($\xi^{\prime}$) and its minor axis to be parallel to the y-axis ($\eta^{\prime}$), we rotate each dSph according to the position angle. 
The radial distance of each star is corrected to the major-axis scale considering the ellipticity.
The background gray-scale density maps and blue and red contours are generated by applying the kernel density estimate (KDE) method smoothed with Silverman’s rule-of-thumb bandwidth \citep{sil86,gou14,sel16}. 
The KDE method can effectively reduce the bias of the binning parameters when constructing histograms. 

In our observed area of Draco, the metal-rich stars (top center panel of Figure~\ref{fig:f_con_kde}) are more centrally concentrated than the metal-poor stars (top left). 
We also find that the density peaks in both panels show a slight offset from the galaxy center, suggesting different formation sites between the metal-poor and metal-rich stars.
The upper right panel shows the cumulative distributions for two RGB groups based on the $hk$ index (solid lines) and the $V-I$ colors (dashed lines). 
We performed a Kolmogorov$-$Smirnov (KS) test to verify whether the distributions of the two groups are different or not. 
The probability value ($p$-value) for the two $hk$-based RGB groups is 0.0003, which is significant at the 0.01 level, and the $p$-value for the $V-I$ based groups is 0.63, which is not significant.

The center panels of Figure~\ref{fig:f_con_kde} show the spatial distributions and the cumulative distributions of the two RGB groups in Sextans dSph. 
Due to the larger size of Sextans ($r$$_{h}$\,=\,27\arcmin) compared to that of Draco ($r$$_{h}$\,=\,10\arcmin), these panels enclose the inner region. 
Albeit weaker than Draco, the distribution of the metal-rich stars in Sextans is slightly more compact than that of the metal-poor stars. 
We note that the metal-poor stars show a curious feature, a strong peak at about 7{\farcm}7 from the center in the minor-axis direction. The peaks of the two populations are apart 8$\arcmin$ from each other.
The possible origin of this off-centered overdensity of the metal-poor stars was discussed as a disrupted star cluster in a separate paper \citep{kim19}. 
The cumulative distributions for Sextans show a weaker separation between the two RGB groups based on $hk$, compared to Draco and CVnI. 
This is mostly due to, as mentioned above, the larger physical size of Sextans than the other two dSphs and thus we observed the relatively inner region. 
To quantify the difference between the RGB groups, we performed the KS test and obtained the $p$-value is 0.080 significant at the 0.1 level. 
The $p$-value for the two groups based on $V-I$ is 0.14, which is less significant.

\begin{figure}[ht!]
\epsscale{1.25}
\plotone{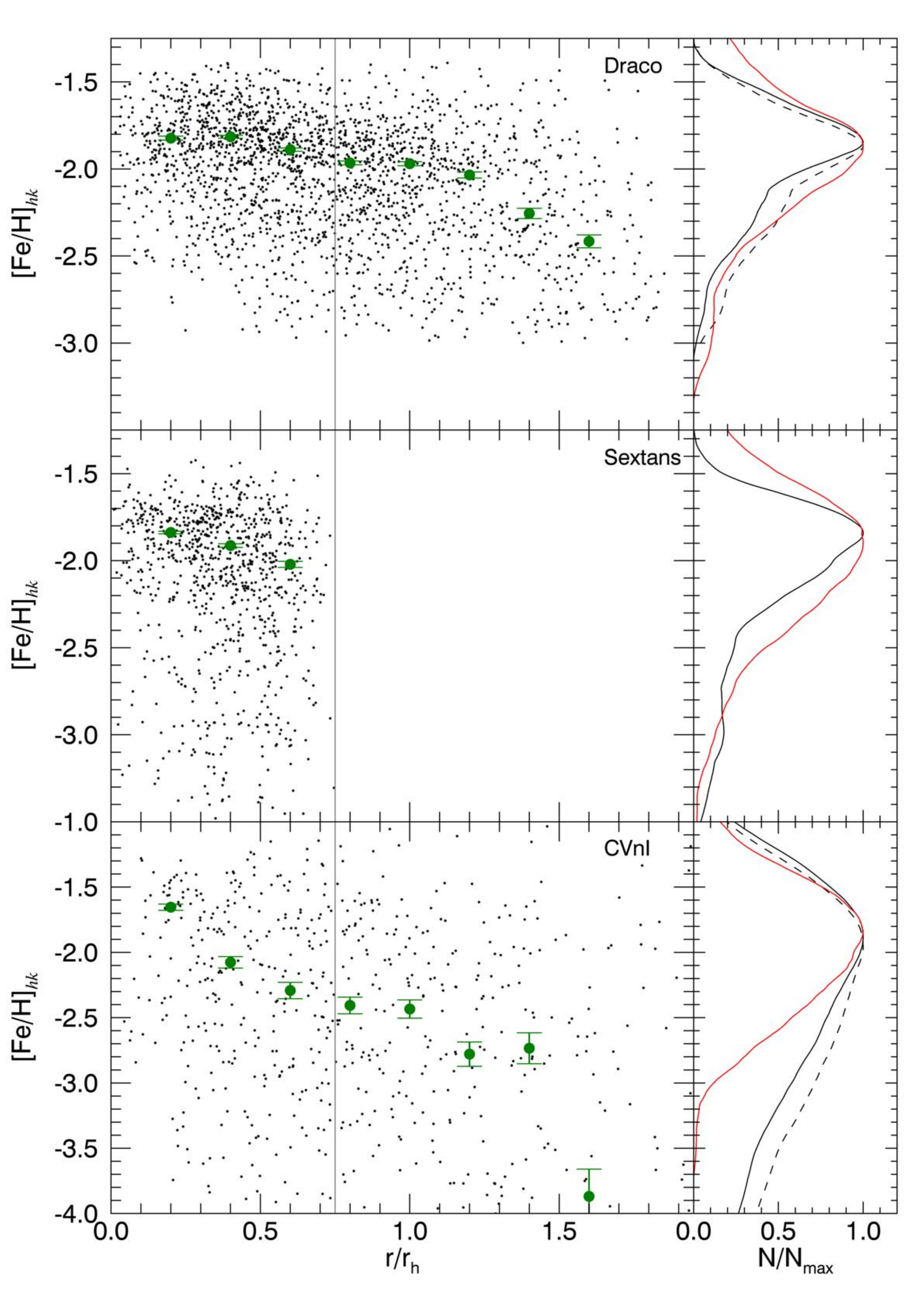}  
\caption{[Fe/H]$_{hk}$ as a function of radius (left) and metallicity distribution functions (right) for Draco (top), Sextans (middle), and CVnI (bottom). Left panels: the green circle and error bars are the mean and the error of the mean ($\sigma$/$\sqrt{N}$) of each radius bin. The gray vertical lines denote $r/r$$_{h}$\,=\,0.75, the approximate observed area of Sextans. Right panels: the solid lines indicate the metallicity distributions for the stars in the inner region ($r/r$$_{h}$ $<$ 0.7), while the dashed lines indicate the distributions for all stars. The red lines represent the spectroscopic metallicity distributions obtained from \citet{kir10}. \label{fig:f_mdf_kde}}
\end{figure}

Same as Draco and Sextans, we plot the spatial distributions of two RGB groups of CVnI in the bottom panels of Figure~\ref{fig:f_con_kde}. 
The metal-rich RGB stars show more concentration at the center than the metal-poor stars, and the highest-density peaks of two groups in the contour maps are separated from each other. 
As well as the contour maps, the cumulative curves indicate the central concentration of the metal-rich RGB stars. 
The $p$-value for the two groups based on $hk$ is 0.019, which is significant at the 0.05 level.
The $p$-value based on $V-I$ is 0.18, consistent with the fact that the $V-I$ color is not a suitable tool for diving metallicity groups.

\subsection{Stellar Radial Distributions and Metallicity Distribution Functions (MDFs)}

The left panels of Figure~\ref{fig:f_mdf_kde} show [Fe/H] of individual stars derived from the $hk$ index ([Fe/H]$_{hk}$) as a function of the projected elliptical radius normalized to the half-light radius ($r/r$$_{h}$).
The [Fe/H]$_{hk}$ values are obtained from [Fe/H]$_{\rm iso}$ 
using the linear correlations between [Fe/H]$_{\rm iso}$ 
and the spectroscopic [Fe/H]$_{ \rm spec}$ (dashed lines in the right panels of Figure~\ref{fig:f_rgb_kirby}).
The mean values of [Fe/H]$_{hk}$ at given radius bins (green circles) decrease with the distance from the center. 
Our $hk$-based investigation finds the negative correlation between the metallicity and galactocentric distance for all three dSphs, whereas previous spectroscopy found a negative correlation for Draco, a positive correlation for Sextans, and no correlation for CVnI \citep{kir11b}.

For Sextans, the distribution of the spectroscopic target stars along the major axis \citep{kir10} would cause the positive correlation. 
As shown in the middle center panel of Figure~\ref{fig:f_con_kde} and Figure 3 of \citet{kim19}, the metal-rich stars distribute along the major axis and this non-isotropic distribution makes the metallicity gradient in the major-axis direction weaker than that in the minor-axis direction.
Hence, the positive slope of Sextans in spectroscopy opposing to our result is most likely due to the sample selection effect together with the smaller sample size.
For CVnI, nearly flat correlation shown from spectroscopy is possibly due to its large distance of the object, which makes the observational errors larger. We speculate that the low brightness of the object combined with the small sample size weakens the negative correlation in CVnI.
The wide-field $VI$ photometry by \citet{oka12} and \citet{oka17} showed that the red HB stars are more concentrated than blue HB stars for Sextants and CVnI, respectively.
If the red HB stars are more metal-rich, this suggests the presence of a negative metallicity gradient, consistent with our results.

On the other hand, the right column of Figure~\ref{fig:f_mdf_kde} shows the metallicity distribution functions (MDFs) for the three dSphs.
The spectroscopic MDFs using data from \citet{kir10} are plotted as red solid lines for comparison. 
The typical dSph MDFs are characterized by a metal-rich peak with a metal-poor tail.
CVnI has a wider distribution than Draco and Sextans due to the larger photometric errors and the steeper correlation between the [Fe/H]$_{\rm iso}$ and [Fe/H]$_{ \rm spec}$ (Figure~\ref{fig:f_rgb_kirby}).
The MDF shape indicates a virtually continuous chemical enrichment with a relatively short timescale.
Indeed, the unimodal, skewed MDFs arise naturally in an aggregate of a large number of protogalactic gas clouds from its virtually continuous chemical evolution through many successive rounds of star formation \citep[e.g.,][]{yoo11,chi14}.

\section{Summary and Discussion}
\label{sec:summarydiscussion}

We have shown that the $hk$ index is correlated with the spectroscopic abundances and can be used as a photometric metallicity indicator. 
The sensitivity of the $hk$ index on the metallicity has been utilized to divide the RGB stars into two groups, metal-poor and metal-rich, to examine the spatial distributions of the different metallicity groups for the three dSphs, Draco, Sextans, and CVnI. 
We have found the presence of the central concentration of metal-rich stars for the three galaxies.
Previous studies found the metallicity gradient in some dSphs using a smaller number of spectroscopic RGB stars and photometric HB stars \citep{har01,tol04,bat06, bat11, kaw06, koc06, kir11b, oka12, oka17}.
Our results confirm the presence of the internal metallicity gradient in the three dSphs, using an even larger number of photometric RGB stars.

To explain the radial metallicity gradient, previous studies suggested several possible mechanisms, such as ram-pressure stripping, gas density, and angular momentum \citep{kir11b,bat14}. 
(1) The ram pressure stripping effectively removes the gas in the outer part that is more loosely bound than the gas that lies inside \citep{sil78}. 
The later-generation stars formed closer to the center from the more metal-rich gas as the gas expelled from outside to inside. 
Thus, this process makes the metallicity gradient with the metal-rich central part. 
(2) The gas density is higher in the center of a galaxy where the gravitational potential well is deeper. 
As gas density is higher, the star formation is increasing \citep{ken83}. 
As a consequence, the metallicity in the central region would be higher than in the outer part.
(3) The higher angular momentum slows down the gas accretion into the galaxy center \citep{schr11}. 
This process keeps the low-metallicity gas in the outer part of the galaxy from flowing into the center. 
Thus, the metal-poor stars were formed at the larger radii.
For (1) and (2) above, the metallicity gradient is interpreted as the inner region experiencing more extended star formation than the outer region. 
For (3), it is vice versa. 
Regardless of the origin of the internal metallicity gradient, the difference in the center positions of the metal-poor and the metal-rich populations supports their different formation epochs.

For a better understanding of the chemical evolution in dSphs, it is needed to use consistent metallicity scales with other studies.
Due to their proxy, the MW clusters, globular and open, have much more abundant spectroscopic data than dSphs in the literature.
With our observational data on some 40 clusters, we are investigating the metallicity effect on the $hk$ index in order to accurately transform our Ca-based photometric metallicity to the spectroscopic metallicity.
Such empirical calibrations of the $hk$ index to various stellar systems, such as monometallic GCs, multimetallic GC, and open clusters \citep[e.g.,][]{hil00,cal07}, are planned in our upcoming papers.

The advantage of the multicolor photometry, the metal-sensitive $hk$ index, and the age-sensitive $V-I$ and $b-y$ colors, combined with the theoretical models, allows better estimation of the age-metallicity relation in dSphs.
In particular, once the metallicities of RGB sequences are fixed based on the metal-sensitive $hk$ index, the age-sensitive $V-I$ and $b-y$ of subgiant branches can be used to determine upper and lower age limits of the oldest and youngest populations, respectively, by comparing population synthesis models. 
We will address this issue in our upcoming papers.

\acknowledgments
This paper is based on data collected at Subaru Telescope, which is operated by the National Astronomical Observatory of Japan. We are grateful to the entire staff at the Subaru Telescope. 
S.-J.Y. acknowledges support by the Mid-career Researcher Program (No. 2019R1A2C3006242) and the SRC Program (the Center for Galaxy Evolution Research; No. 2017R1A5A1070354) through the National Research Foundation of Korea.
N.A. is supported by the Brain Pool Program, which is funded by the Ministry of Science and ICT through the National Research Foundation of Korea (2018H1D3A2000902).

%

\vspace{5mm}
\facilities{SUBARU (Supreme-Cam)}





\end{document}